\documentclass[runningheads,a4paper]{llncs}

\usepackage{amsmath}
\usepackage[xcdraw]{xcolor}
\usepackage{amssymb}
\usepackage{graphicx}
\usepackage{epstopdf}
\usepackage[hyphens]{url}
\usepackage{cite}
\usepackage{geometry}
\usepackage{booktabs}

\usepackage[flushleft]{threeparttable}

\usepackage{arydshln}
\usepackage{multirow}

\usepackage{pifont}
\newcommand{\cmark}{\ding{51}}%
\newcommand{\xmark}{\ding{55}}%

\usepackage{listings, color}

\definecolor{gray}{rgb}{0.4,0.4,0.4}
\definecolor{darkblue}{rgb}{0.0,0.0,0.6}
\definecolor{cyan}{rgb}{0.0,0.6,0.6}

\lstdefinestyle{listXML}{language=XML, extendedchars=true,  belowcaptionskip=5pt, xleftmargin=1.8em, xrightmargin=0.5em, numbers=left, numberstyle=\small\ttfamily\bf, frame=single, breaklines=true, breakatwhitespace=true, breakindent=0pt, emph={}, emphstyle=\color{red}, basicstyle=\small\ttfamily, columns=fullflexible, showstringspaces=false, commentstyle=\color{gray}\upshape,
	morestring=[b]",
	morecomment=[s]{<?}{?>},
	morecomment=[s][\color{orange}]{<!--}{-->},
	keywordstyle=\color{cyan},
	stringstyle=\color{black},
	tagstyle=\color{darkblue},
	morekeywords={xmlns,version,type}
}

\lstdefinestyle{customc}{
	belowcaptionskip=1\baselineskip,
	belowcaptionskip=5pt, xleftmargin=1.8em, xrightmargin=0.5em, numbers=left, numberstyle=\small\ttfamily\bf,
	breaklines=true, breakatwhitespace=true, breakindent=0pt, emph={}, emphstyle=\color{red}, basicstyle=\small\ttfamily, columns=fullflexible, showstringspaces=false, commentstyle=\color{gray}\upshape,
	breaklines=true,
	extendedchars=true,
	frame=single,
	morecomment=[s][\color{orange}]{/*}{*/},
	xleftmargin=\parindent,
	language=Java,
	showstringspaces=false,
	basicstyle=\footnotesize\ttfamily,
	keywordstyle=\bfseries\color{green!40!black},
	commentstyle=\itshape\color{purple!40!black},
	stringstyle=\color{orange},
	keywordstyle=\color{cyan},
	stringstyle=\color{black},
	tagstyle=\color{darkblue},
	morekeywords={xmlns,version,type}
}

\begin{document}

\title{When Theory and Reality Collide:  Demystifying the Effectiveness of Ambient Sensing for NFC-based Proximity Detection by Applying Relay Attack Data}

\author{Iakovos Gurulian \and Carlton Shepherd \and
	Konstantinos Markantonakis \and Raja Naeem Akram \and  Keith Mayes}

 \institute{Information Security Group, Smart Card Centre, Royal Holloway, University of London. United Kingdom,\\
}

\maketitle

\begin{abstract}
Over the past decade, smartphones have become the point of convergence for many applications and services. There is a growing trend in which traditional smart-card based services like banking, transport and access control are being provisioned through smartphones. Smartphones with Near Field Communication (NFC) capability can emulate a contactless smart card; popular examples of such services include Google Pay and Apple Pay. Similar to contactless smart cards, NFC-based smartphone transactions are susceptible to relay attacks. For contactless smart cards, distance-bounding protocols are proposed to counter such attacks; for NFC-based smartphone transactions, ambient sensors have been proposed as potential countermeasures. In this study, we have empirically evaluated the suitability of ambient sensors as a proximity detection mechanism for contactless transactions. To provide a comprehensive analysis, we also collected relay attack data to ascertain whether ambient sensors are able to thwart such attacks effectively. We initially evaluated 17 sensors before selecting 7 sensors for in-depth analysis based on their effectiveness as potential proximity detection mechanisms within the constraints of a contactless transaction scenario.  Each sensor was used to record 1000 legitimate and relay (illegitimate) contactless transactions at four different physical locations. The analysis of these transactions provides an empirical foundation on which to determine whether ambient sensors provide a strong proximity detection mechanism for security-sensitive applications like banking, transport and high-security access control.  

\end{abstract}

\section{Introduction}\label{Introduction}
Near Field Communication (NFC) \cite{Coskun2013} has opened smartphone platforms to many different application domains, particularly smart cards. By enabling a smartphone to emulate a contactless smart card, users may use a mobile device as a potential replacement for smart cards in applications such as banking, transport and access control.   Leading technology firms, particularly Google and Apple, have already launched smartphone-based payment services.  To this end, the Android platform has proposed Host-based Card Emulation (HCE) \cite{umar2015performance} that is poised to open up card emulation via NFC to any application on an Android smartphone.

Mobile (contactless) payments are being adopted by tech-savvy, young age groups \cite{UKCardsPayment2015a}.  Deloitte predicted that 5\% of the 600-650 million NFC-enabled mobile phones would be used at least once a month in 2015\footnote{Actual figures for 2015 were not available at the time of writing this paper. However, early forecasts for the global market in mobile payments for 2016 and beyond are available at \url{http://www.nfcworld.com/technology/forecast/}} to make a contactless payment \cite{Deloitte2015}.  We can reasonably assume, therefore, that mobile payments will be a significant payment medium in the future, potentially overtaking contactless smart cards. According to Statista, in 2015, 12.7\% of smartphone users in the USA were actively using proximity mobile payment and  the value of such transactions is projected to grow to 114 billion US\$ by 2018\footnote{Statista website: \url{http://www.statista.com/statistics/244475/proximity-mobile-payment-transaction-value-in-the-united-states/}}. Although this discussion is limited to the financial industry, similar trends are being observed in other domains where contactless smart cards are deployed \cite{VeriFone2010}.

A relay attack is a passive man-in-the-middle attack in which an attacker extends the distance between a genuine payment terminal (point-of-service) and a genuine contactless smart card (or NFC-enabled mobile device). This attack can enable a malicious user to access services for which the genuine user is eligible, such as paying for goods or accessing a building with physical access controls. Smart card and smartphone-based contactless transactions are susceptible to relay attacks. Such attacks on contactless smart cards have been extensively studied in the literature \cite{DrimerM07,Francillon11,Hancke2009615,kfir2005picking}. With the advent of NFC technology to emulate a smart card, relay attacks were demonstrated to be successful in \cite{Francis:2010:PNP:1926325.1926331,FrancisHMM11,5741305}. To counter such attacks, ambient sensors available on smartphones are being proposed \cite{Halevi2012,6378376,shrestha2014drone,truong2014comparing,Urien201428,Varshavsky2007} as strong candidates. 

In this investigation, we carried out an empirical study that included both legitimate and illegitimate transaction data. Legitimate data was collected from two devices that were in close proximity ($<$3cm) to each other, whereas illegitimate data was collected from two devices that were 5ft (1.5m) apart. This provided real data, collected from our field trials, to analyse ambient sensor data for devices that are both in proximity to each other and at a distance.  The primary contributions of this paper include:
\begin{itemize}
\item{Evaluation Test-bed:} We established a reproducible\footnote{Test-bed implementation and data collected from the field trials would be made public on project website \url{anonymous}} test-bed environment that was used to collect field data for legitimate and illegitimate transactions concurrently at a single point in time `$Time_i$'. For detailed discussion of the theoretical model and its practical implementation, see section \ref{sec:Evaluation-Framework-for-Single-Sensor-Single-Tap}.
\item{Proximity Detection Analysis:}
The suitability of a proximity detection mechanism for critical applications, such as banking, transport and high-security access control, is based on its ability to uniquely pair measurements taken from a payment terminal and a payment instrument (or mobile handset in this case). This is to provide confidence that the two devices were truly in close proximity ($\approx$ 3cm) to each other. The measurements should be taken within 500ms as mandated by the EMV specifications that govern contactless transactions for the banking sector \cite{VISAMobileTicketing2013,MasterCard2014,MasterCardTaB,EVM2015-ContactlessArchitectureReq,VISA-TADG}. Section \ref{sec:CalculatingFPRFNRandEER} discusses the analysis results for both the proximity detection and effectiveness analysis. 
\item{Effectiveness Analysis Against Relay Attacks:}
The effectiveness of an ambient sensor in countering a relay attack was studied using the illegitimate transaction data. An effective ambient sensor should be able to reject sensor values that were recorded from two devices further than 3cm from each other, while still accepting all legitimate transactions. In our experiments, we selected a distance of 5ft (1.5m) between the devices; at this realistic distance the ambient sensor should be able to perform well as anti-relay countermeasure. 
\end{itemize}

\section{Ambient Sensors in Mobile-NFC Transactions}
\label{AmbientSensorinMobile-NFCTransaction}
In this section, we briefly discuss NFC-based smartphone transactions and how relay attacks are carried out. Subsequently, we discuss ambient sensor deployment to avoid such attacks. 

\subsection{Relay Attacks on Contactless Transactions}
\label{sec:Relay-Attacks-on-Contactless-Transactions}
In an NFC-based mobile contactless transaction, a mobile handset is brought into the radio frequency range ($<$3cm) of a payment terminal through which it can initiate a dialogue.  During a contactless transaction, physical contact is not necessary and, in many cases, a second factor of authentication, e.g.\ biometrics or Personal Identification Number (PIN), is not required \cite{EVM2015-ContactlessArchitectureReq}. This makes it difficult to ascertain whether the genuine or relay device is in close proximity of the terminal. It should be noted, however, that the use of a PIN or biometric may not counter a relay attack effectively in certain situations (notably the Mafia fraud attack \cite{Cremers2012}).

\begin{figure}[ht]
	\centering
		\includegraphics[width=0.65\columnwidth]{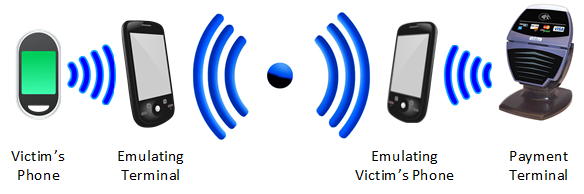}
	\caption{Overview of a Relay Attack}
	\label{fig:relay_attack}
\end{figure}

In a relay attack \cite{Francis:2010:PNP:1926325.1926331,FrancisHMM11,5741305}, shown in Figure \ref{fig:relay_attack}, an attacker must present a malicious payment terminal to a genuine user and a masquerading payment instrument (mobile phone) to a genuine payment terminal.  The goal of the malicious actor is to extend the physical distance of the communication channel between the victim's mobile phone and the payment terminal -- relaying each message across this extended distance.  The attacker has the potential to gain access to services using the victim's account if it successfully relays messages without detection.

\subsection{Ambient Sensors: Single Sensor, Single Tap Proximity Detection}
\label{sec:Ambient-Sensors-Single-Sensor-Single-Tap-Proximity-Detection}


A substantial portion of the work surrounding relay-attack countermeasures for contactless smart cards relates to distance bounding protocols \cite{rasmussen2010realization,DrimerM07,Francillon11,Hancke:2008:ATD:1352533.1352566,Cremers2012,boureanu2014towards}.  However, these may not be feasible for NFC-enabled phones -- at the current state of the art -- due to their requirement of high time-delay sensitivity and specialised hardware \cite{Coskun2013,Halevi2012}.  Thus, several methods have been proposed to provide some notion of proximity detection for NFC-based contactless mobile transactions, most of which use environmental (ambient) sensors present on modern mobile handsets (for related work see Section \ref{sec:RelatedWork}).  In the following section, we discuss how ambient sensors have been proposed to counter relay attacks in NFC-based mobile contactless transactions. 

\begin{figure}[ht]
	\centering
		\includegraphics[width=0.8\textwidth]{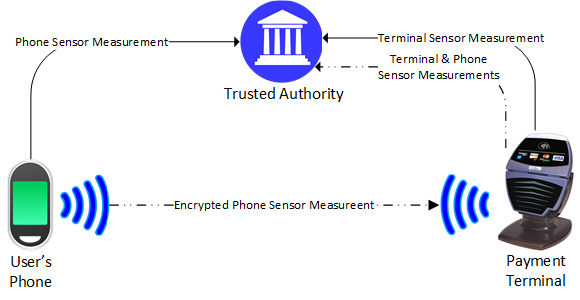}
	\caption{Generic Deployment of Ambient Sensors as Proximity Detection Mechanisms}
	\label{fig:AmbientSensorDeploymentOverview}
\end{figure}

An ambient sensor measures a particular physical/environmental property of its immediate surroundings, such as temperature, light or sound; modern smartphones and tablets are equipped with one or more of these sensors. The physical environment surrounding a smartphone (or a payment terminal) can potentially provide a rich set of attributes that may be unique to that location -- the sound and lighting of a quiet, brightly-lit room, for example -- and such information may be crucial for proximity detection between two interacting devices.  

In this section, we discuss a generic approach for deploying ambient sensing as a proximity detection mechanism that can be used in the context of mobile payments. Figure \ref{fig:AmbientSensorDeploymentOverview} illustrates the entities involved in this process.  Variations of this approach are discussed below:
\begin{enumerate}
\item \textbf{Independent Reporting}.  In this scenario, depicted as solid lines in Figure \ref{fig:AmbientSensorDeploymentOverview}, both the smartphone and payment terminal collect sensor measurements independently of each other and transmit these to a trusted authority.  This authority compares the sensor measurements, based on some predefined comparison algorithm with set margins of error (threshold), and decides whether the two devices are in proximity to each other.
\item \textbf{Payment Terminal Dependent Reporting}.  This setup, depicted as double-dot-dash lines in Figure \ref{fig:AmbientSensorDeploymentOverview}, involves the smartphone encrypting the sensor measurements with a shared key (between smart phone and trusted authority) and transmitting the encrypted message to the payment terminal.  The payment terminal sends its own measurements and the smartphone's to the trusted authority for comparison.
\item \textbf{Payment Terminal (Localised) Evaluation}.  The smartphone transmits its own measurement to the payment terminal, which then compares it with its own measurements locally; the payment terminal then decides whether the smartphone is in close proximity. 
\end{enumerate}

Regardless of how the user interacts with the payment terminal, e.g.\ touching or tapping it with their device, the overall deployment architecture falls under one of the above scenarios.  It can be observed that there is a potential for a fourth scenario in which smartphone (payment instrument) could perform the comparison.  We assume in this paper, however, that smartphones might be in control of malicious users due to which we ignored this scenario.  In this study, our sole focus is on a single tap-single sensor deployment; the evaluation framework of such a deployment is discussed in subsequent sections. 

\section{Evaluation Framework for Single Sensor -- Single Tap}
\label{sec:Evaluation-Framework-for-Single-Sensor-Single-Tap}
In this section, we discuss the theoretical evaluation framework and provide the rationale behind the setup of the data collection platform (part of the Evaluation Test-bed). We describe how data was collected and analysed in subsequent sections. 


\subsection{Evaluation Framework: A Theoretical Model}
\label{sec:Evaluation-Framework-A-Theoretical-Model}
As mentioned previously, we aim to analyse the effectiveness of ambient sensors, not only in proximity detection but also in their ability to counter relay attacks. Towards this goal, a test environment must be devised in which we can collect both legitimate and illegitimate pairs of ambient sensor measurements. A genuine pair of sensor measurements are from two devices that are physically in close proximity to each other ($<$3cm). An illegitimate pair is from two devices that are not physically in close proximity, and in the context of this paper we consider this distance to be 5ft (1.5m).

To achieve this, we combine a genuine transaction framework, as shown in Figure \ref{fig:relay_attack}, and a relay attack framework (Figure \ref{fig:AmbientSensorDeploymentOverview}) in such a way that we have both the genuine pair and illegitimate pair measurements at some transaction time interval `$Time_i$'. The constraint of time to `$Time_i$' is imposed because ambient sensor values may change depending upon the location and time at which the transaction was carried out. To avoid any discrepancies introduced because of the dependence of ambient sensors on location and time, we collected these pairs concurrently. This also helps us understand how different the sensor measurements would be at the time of transaction `$Time_i$', from a close proximity device ($<$3cm) and distant device (5ft).

\begin{figure}[ht]
	\centering
		\includegraphics[width=0.85\textwidth]{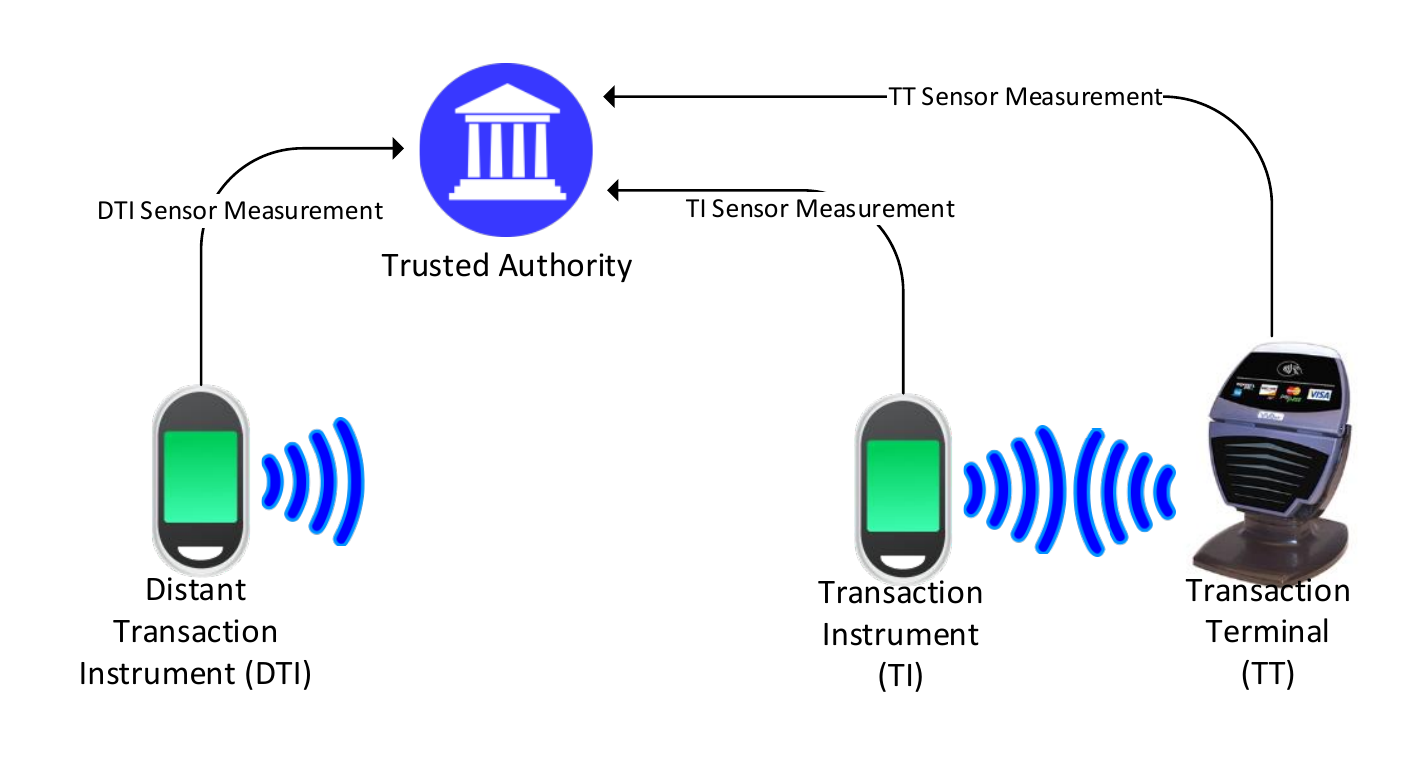}
	\caption{Overview of Test-bed  -- Trial Data Collection Platform}
	\label{fig:TrialDataCollectionPlatform}
\end{figure}

Figure \ref{fig:TrialDataCollectionPlatform} shows our data collection setup. The Transaction Terminal (TT) is a static device, and we use this as a reference point for our two pairs. The Transaction Instrument (TI) is a mobile phone in close proximity to the TT at time `$Time_i$' and the ambient sensor measurement pair TT-TI is referred to as the genuine pair. Another mobile phone, at a 5ft (1.5m) distance from the TT, is referred to as the Distant Transaction Instrument (DTI). The ambient sensor measurement pair TT-DTI taken at time `$Time_i$' is referred to as the illegitimate pair.  The rationale for setting up the test environment in this manner is to collect ambient sensor values from proximate and distant devices simultaneously. If, for a transaction `$T_i$', the genuine pair is uniquely identified (and accepted) then the ambient sensor is considered effective. However, if the illegitimate pair is accepted (whether uniquely or along with the genuine pair) then the relay attack on that `$T_i$' is successful. The reasoning is that if two devices at 5ft (1.5m) apart measure ambient sensors independently, and the illegitimate pair is indistinguishable from a genuine pair, then the attacker can successfully relay messages between these two devices without being detected. 

At a point in time, the beginning of $Time_i$, a user taps TI to TT; at this point, DTI also initiates the ambient sensor measurements along with TI and TT.  Thus, at $Time_i$, we have three separate ambient sensor measurements for the three devices.  We acquired the ground truth of genuine and illegitimate pairs as the transactions were conducted with the help of volunteers from the university in a monitored environment.  Overall, for an ambient sensor to be effective, the Trusted Authority should be able to distinguish the genuine pair from the illegitimate pair. To evaluate each ambient sensors’ effectiveness for proximity detection and to detect relay attacks, we analysed the collected data using two methods:
\subsubsection{Evaluation Method 1 - Proximity Detection}
In this analysis, we considered only the genuine pairs (TT-TI) of ambient sensor measurements, based on this empirically collected data from field trials at four different locations on the university campus. We calculated the threshold, $t$, which would then be used by entities such as the Trusted Authority (in Figure \ref{fig:TrialDataCollectionPlatform}) to make the decision about whether devices participating in a transaction are proximate to each other or not. Based on this $t$, we evaluated how many of the transactions represented by genuine pairs would be accepted or not. This analysis provides quantitative results demonstrating the accuracy of ambient sensors for proximity detection using only TT-TI pairs.

\subsubsection{Evaluation Method 2 - Inclusion of Attack Vector}
Next, we included the ambient sensor measurements taken by DTI -- a device that is 5ft (1.5m) away from the TT. We then recomputed these $t$ values (from Evaluation Method 1) using data included from DTI, and evaluated whether the illegitimate pair (TT-DTI) would be accepted as a genuine pair, while considering TT-TI pairs as legitimate. This analysis provided the success and failure rates of a potential relay attack for a particular ambient sensor, while also judging the rate at which legitimate transactions are rejected/accepted.

\subsection{Test-bed Architecture}
\label{sec:Test-bed-Architecture}

As mentioned in Section~\ref{sec:Evaluation-Framework-A-Theoretical-Model}, three devices were used in the data collection phase, TT, TI and DTI\@.  A distinct application was developed to assist in data collection, which was then installed on each of the three devices.  During the experimental phase, the devices TT and DTI were placed on tables at a distance of 5ft (1.5m) from one another.  Both devices (TT and DTI) were placed on stands facing the same direction.

\begin{figure}
	\centering
	\includegraphics[width=0.9\textwidth]{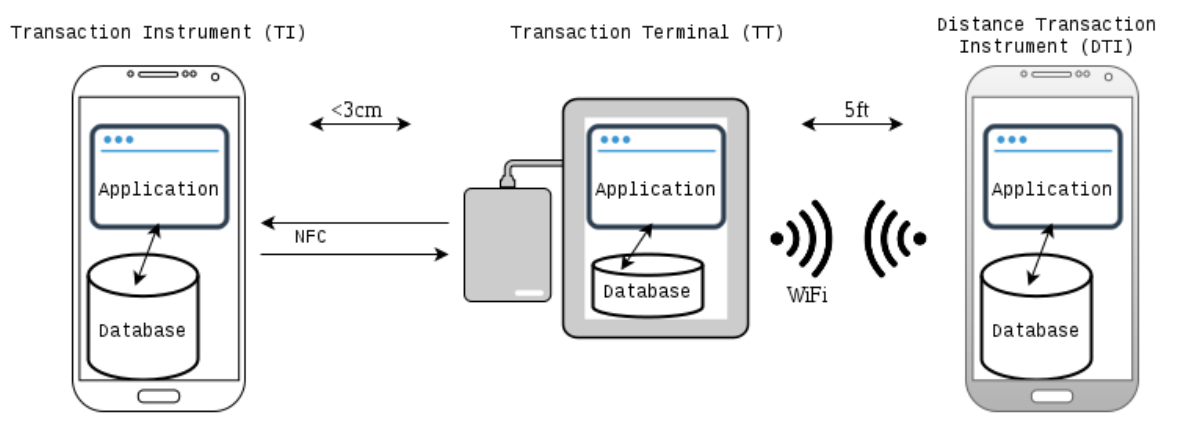}
	\caption{Test-bed Architecture}
	\label{fig:architecture}
\end{figure}

When TI was brought in close proximity to TT, an NFC connection between the two devices would be established, initiated by TT, indicating the beginning of a transaction.
According to the EMV standard, TT and TI should be in proximity, less than 3cm apart~\cite{EVM2015-ContactlessArchitectureReq}.
Device TT would then immediately broadcast a message on the local network, to be received by device DTI, notifying it of the transaction initiation process.
All three devices would then record values from a predefined ambient sensor for 500ms and then store it in a local database file.
Figure~\ref{fig:architecture} depicts the test-bed architecture.

\subsection{Data Collection Implementation Platform and Specifications}

Three Samsung Galaxy S4 (GT-I9500) Android devices running on Android 5.0.1 were used in the experimental phase, with all extraneous synchronisation services and applications disabled. This specific device was found to include a large variety of sensors, covering the majority of sensors supported by the Android platform~\cite{androidsensors}, excluding the Geomagnetic Rotation Vector.
An overview of each sensor is provided in Appendix~\ref{sec:AmbientSensors}.

Table~\ref{tab:availability} lists the available sensors on the device, the ones used in our experiments and the rational behind excluding some.
For the majority of excluded sensors, no values could be captured in the 500ms timeframe.
With the exception of the proximity sensor, additional time is required for these sensors in order to return data.
According to~\cite{Anonymous}, less than 5\% of the time any data is being returned by the Bluetooth, GPS, Network Location and WiFi sensors.
Additionally, we evaluated the Ambient Temperature and Relative Humidity sensors, and discovered that data was also returned less than 5\% of the time when recorded for 500ms, for 50 repetitions.
The proximity sensor on Samsung Galaxy S4 returns only a true or false value when the sensor, located in the front of the device, is covered or uncovered, hence it could not be used effectively in the experimental phase.
Lastly, the sound sensor, i.e.\ the microphone, of the specific device was tested; we found that this could not initiate and record values within 500ms. For our analysis we used the 7 best sensors and discarded the others. 

\begin{table}[ht]
	\centering
	\caption{Sensor Availability for Samsung Galaxy S4}
	\label{tab:availability}
		\begin{threeparttable}
			\begin{tabular}{@{}lc@{\hspace{0.25cm}}cl@{}}
				\toprule
				\multicolumn{1}{c}{\textbf{Sensors}} & \multicolumn{1}{c@{\hspace{0.25cm}}}{\textbf{Supported}} & \multicolumn{1}{c}{\textbf{Used}} & \multicolumn{1}{c}{\textbf{Reason}} \\ \midrule
				\textbf{Accelerometer} & \cmark & \cmark & \-- \\
                \textbf{Gravity} & \cmark & \cmark & \-- \\
				\textbf{Gyroscope} & \cmark & \cmark & \-- \\
                \textbf{Light} & \cmark & \cmark & \-- \\
				\textbf{Linear Acceleration} & \cmark & \cmark & \-- \\
				\textbf{Magnetic Field} & \cmark & \cmark & \-- \\
                \textbf{Rotation Vector} & \cmark & \cmark & \-- \\
				\textbf{Ambient Temperature} & \cmark & \xmark & Not enough values in timeframe. \\ 
				\textbf{Bluetooth} & \cmark & \xmark & Not enough values in timeframe. \\
				\textbf{GPS} & \cmark & \xmark & Not enough values in timeframe. \\
				\textbf{Relative Humidity} & \cmark & \xmark & Not enough values in timeframe. \\
				\textbf{Network Location} & \cmark & \xmark & Not enough values in timeframe. \\
				\textbf{Pressure} & \cmark & \xmark & Not mature for data collection. \\
				\textbf{Proximity} & \cmark & \xmark & Not enough values in timeframe. \\
				\textbf{Sound} & \cmark & \xmark & Not enough values in timeframe. \\
				\textbf{WiFi} & \cmark & \xmark & Not enough values in timeframe. \\
				\textbf{GRV$^\dag$} & \xmark & \xmark & Not present. Used Rotation Vector instead. \\ \bottomrule
			\end{tabular}
			\dag Geomagnetic Rotation Vector
		\end{threeparttable}
\end{table}

\subsubsection{Android applications}

Three Android applications were developed for TT, TI and DTI.
The application built for device TT included two connection interfaces. 
The first interface, used for the NFC connection with the device TI, device TT was set to NFC reader mode, allowing it to interact with discovered NFC tags.
The second, used for connection with the device DTI, device TT would broadcast a UDP packet in the local network, setting the destination port to 8888.
The application running on device TI, in order to achieve NFC connection with TT, was using Host-based Card Emulation (HCE)~\cite{androidhce}.
HCE allows Android-powered devices to act as contactless smart-cards and interact with NFC readers.
Finally, the application running on device DTI featured a broadcast listener, listening on network port 8888 for UDP packets from TT\@.
The source code of the three applications is available at \url{https://anonymised.com/}.

When TI was brought into proximity with TT, the former would send a message to the first via NFC\@.
This message consisted by a random 7-byte transaction ID, generated by the device TT and used to uniquely identify each transaction during the analysis phase, and the sensor to be recorded in the current transaction. Following the transmission of the message to TI, device TT would broadcast the same message to the wireless local area network.
Devices TT and DTI were connected to the same wireless hotspot, created for the requirements of the experiment.
No other devices were connected on the hotspot.
The device DTI would then capture the message by listening for broadcast UDP packets with destination port 8888.  The two devices, TI and DTI, would start recording data for a period of 500ms, using the sensor specified in the message that they received from TT, upon receiving it.
Device TT would initiate the recording process, for the same amount of time and using the same sensor immediately after sending the two messages.
This is done so that the three devices initiate the recording process as close to each other as possible.

On the Android operating system, data captured by a sensor are returned to an application in time intervals set by the application.
The rate at which data was polled from the sensors was set at the highest available rate supported by the devices.
The highest available rate was the same across all three devices, as they shared the same hardware and software.
By default, the operating system only returns data values when the values alter.

Following the recording time period, device TI would send a response message, containing the transaction ID and sensor, to TT\@.
The former would then validate that the two devices were on the same transaction and recording the same sensor, and in case of inconsistencies would not store data for the specific transaction.
After validation, the three devices would then store the recorded data in a local SQLite database, along with the transaction ID, sequence number, timestamp and the pre-defined location in which the recording took place.
Separate database tables were used for each sensor.
The recorded data was stored in XML format.
During the data analysis phase, only transactions that existed in all three databases, based on their transaction ID were considered.

\begin{figure}
	\centering
	\includegraphics[width=0.9\textwidth]{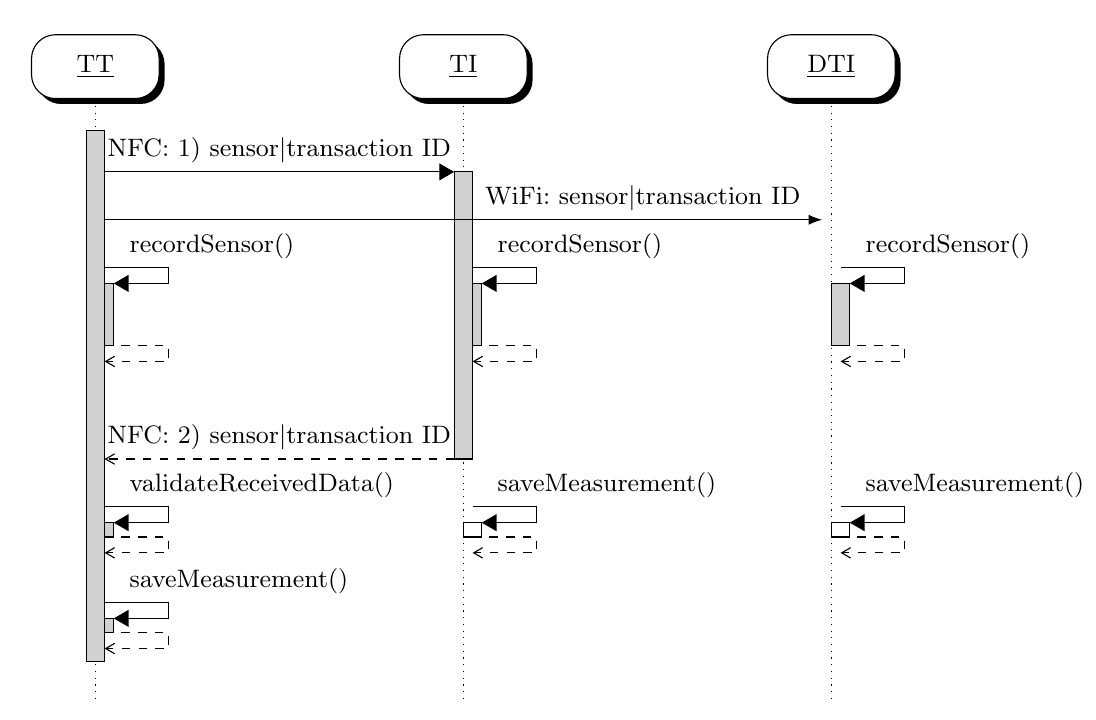}
	\caption{Measurement Recording Overview}
	\label{fig:diagram}
\end{figure}

An overview of the measurement recording across the three devices is presented in Figure~\ref{fig:diagram}.
After the recording of a measurement, device TT would automatically switch to the next sensor in the list of examined sensors and wait for a new transaction to be initiated.

%
%
%


For the data collection, student volunteers were requested to participate in the trails. 
A total of 1000 transactions were collected for each sensor, distributed in 4 distinct locations on the university campus.
Individual volunteer was asked to tap the device TI to TT, while at the same time another volunteer was tapping a dummy device to DTI.

\section{Evaluating Ambient Sensors for Proximity Detection}

In line with the framework discussed previously, we conducted a two-fold evaluation using the sensor data collected during field trials:
\begin{itemize}
	\item \textbf{Evaluation Method 1}: The success/failure rates for correctly identifying the two legitimate devices in a transaction were determined (transaction instrument, $TI$, and terminal, $TT$).  That is, given a set of transactions $\textbf{T}$, with each transaction comprising a triple of TI, terminal and distant instrument ($DTI$) sensor measurements, i.e. $\textbf{T}_{i} = (TI_{i}, TT_{i}, DTI_{i})$, can $TI_{i}$ and $TT_{i}$ be identified as belonging to transaction $i$.  All $TI$ values were compared with with all values of $TT$ in a $\lvert\textbf{T}\rvert \times \lvert \textbf{T}\rvert$ matrix.  We aim to generate a threshold, $t$, that accepts all valid pairs, i.e. $(TI_{i}, TT_{j}), i=j$, while rejecting those which are invalid: $(TI_{i}, TT_{j}), i \neq j$.
		\\
	\item \textbf{Evaluation Method 2}: The rates of the distant device being accepted as legitimately co-located with the terminal, i.e. whether $DTI_{i}$ is accepted incorrectly as a legitimate TI measurement with the terminal $TT_{i}$ despite being 5ft (1.5m) away.  This way, the success/failure rate of a relay attack being conducted successfully can be determined in the presence of ambient sensing.  We generate new thresholds that maximise the number of legitimate transactions, i.e. $(TI_{i}, TT_{i})$, while minimising the number of accepted transactions conducted from afar between $(DTI_{i}, TT_{i})$.
\end{itemize}

\subsection{Transaction Data Analysis}
\label{sec:sim}

In this section, we describe how the data analysis was conducted. 

\subsubsection{Sensor Measurement Similarity}
To determine whether the sensor measurements of $TI_{i}$ and $DTI_{i}$ were in proximity with $TT_{i}$, two metrics were employed to judge the similarity between, firstly, $TI_{i}$ and $TT_{i}$ for Evaluation 1, and $(TI_{i}, TT_{i})$ and $(DTTI_{i}, T_{i})$ for Evaluation 2.  Specifically, we used the Mean Absolute Error and Pearson's Correlation Coefficient, as used in \cite{mehrnezhad2014tap}, which are given in Eqs.~\ref{eq:mae} and \ref{eq:corr} respectively.

\begin{equation}
	MAE(A_{i}, B_{i}) = \frac{1}{N}\sum_{j=0}^{N} | A_{i,j} - B_{i,j} |
	\label{eq:mae}
\end{equation}
Where $N$ refers to the number of datapoints in a sensor measurement, and $A_{i,j}$ refers to the $j^{th}$ datapoint of the $i^{th}$ sensor measurement on device $A$.
\begin{equation}
	corr(A_{i}, B_{i}) = \frac{cov(A_{i}, B_{i})}{\sigma_{A_{i}} \cdot \sigma_{B_{i}}}
	\label{eq:corr}
\end{equation}
Where $\sigma_{A_{i}}$ denotes the standard deviation of the sensor measurements of $A_{i}$, and $cov$ represents the covariance, given below in Eq. \ref{eq:cov}, with $\mu_{A_{i}}$ denoting the mean of $A_{i}$.
\begin{equation}
	cov(A_{i}, B_{i}) = \frac{1}{N} \sum^{N}_{j=0}(A_{i,j} - \mu_{A_{i}})(B_{i,j} - \mu_{B_{i}})
	\label{eq:cov}
\end{equation}
\begin{equation}
	M = \sqrt{x^{2} + y^{2} + z^{2}}
	\label{mageq}
\end{equation}

\subsubsection{Pre-processing}

All sensors except those for light produce a vector of values consisting of $x$, $y$ and $z$ components.  For these sensors, the vector magnitude (Eq.\ \ref{mageq}) was used as a general-purpose method for producing a single, combined value prior to computing the MAE and correlation coefficient.  In the event that the sensor values exceeded the maximum permitted transaction time (500ms), such values were discarded prior to computing the similarity.  Moreover, any sensor values on device $A$ that were recorded after the maximum time recorded by device $B$ were also discarded.  This was to prevent undefined results when computing MAE; calculating $|A_{i,j} - B_{i,j}|$ is undefined when $A_{i,j}$ has no corresponding datapoint on device $B$, i.e. $B_{i,j}$.  

\begin{figure*}
	\centering
	\includegraphics[width=\columnwidth]{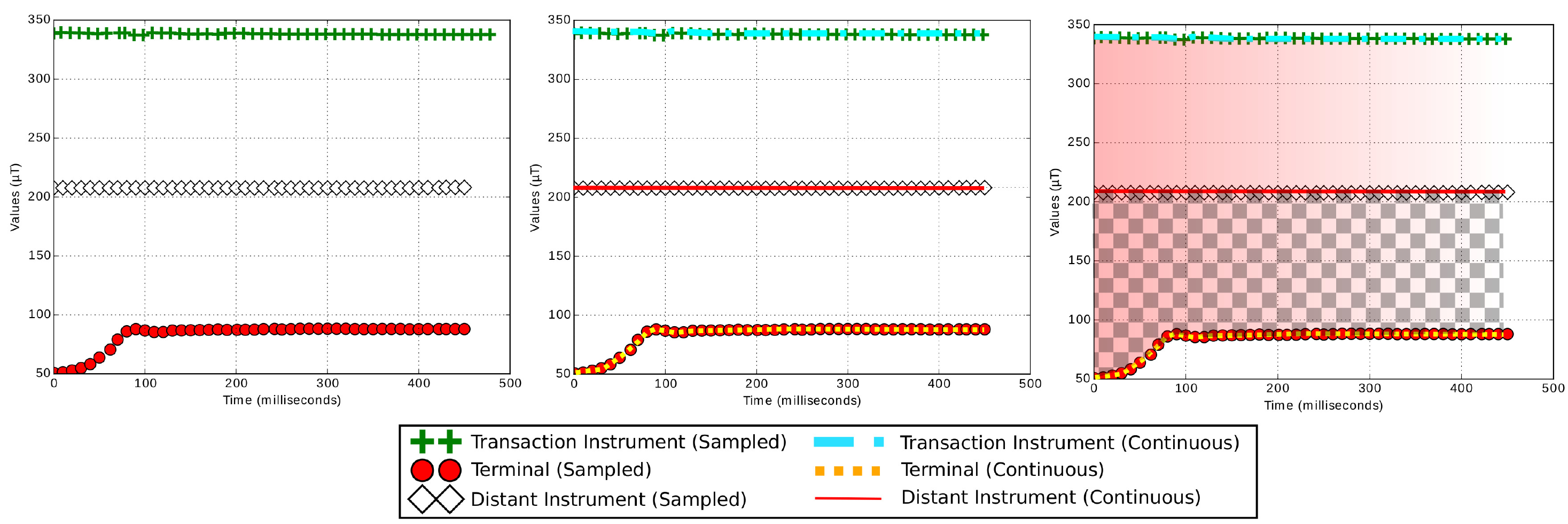}
	\caption{Pre-processing and similarity workflow for a Magnetic Field-based transaction using MAE (left to right): \textbf{(a)} Sensor data acquisition from all three devices, \textbf{(b)} Perform linear interpolation and truncation, \textbf{(c)} Measure MAE between terminal and distant device (shaded area) and terminal and transaction instrument (checkered region).}    \label{fig:workflow}
\end{figure*}

Related to this was the issue of irregular sampling rates; usually, sensor values of device $A$ did not map to exactly the same times as those recorded on $B$.  That is, $A_{i,3}$ may have been collected at 15ms from the start, while $B_{i,3}$ was recorded at 22ms.  This time discrepancy will have some impact on the results, particularly if the number of recorded datapoints is small and the time difference is large.  Hence, to mitigate this, linear interpolation was performed on both sets of datapoints to establish a consistent sampling rate of 10ms before computing either similarity metric.  We summarise the data analysis workflow in Figure \ref{fig:workflow}.

\subsection{Calculating FPR, FNR and EER}
\label{sec:CalculatingFPRFNRandEER}

To perform Evaluation 1 and 2, we computed Equal Error Rates (EERs) to determine optimal similarities/thresholds where the rate of false positives (FPR) is equal to the rate of false negatives for each tested sensor.  The following notions of true positives (TPs) and negatives (TNs), false positives (FPs) and negatives (FNs) are defined for each evaluation:
\begin{itemize}
	\item \textbf{Evaluation 1}. TP: $TI_{i}$ is paired with the correct, corresponding $TT_{i}$.  TN: $TI_{i}$ and $TT_{j}$ ($i \neq j$) is rejected correctly.  FP: $TI_{i}$ and $TT_{j}$ is accepted as belonging to the same transaction.  FN: $TI_{i}$ and $TT_{j}$ is rejected, despite belonging to the same transaction.\\

	\item \textbf{Evaluation 2}.  TP: the legitimate transaction instrument-terminal pair ($TI_{i}, TT_{i}$) is accepted correctly.  TN: the distant instrument-terminal pair, ($DTI_{i}, TT_{i}$), is rejected correctly.  FP: the distant-terminal pair is wrongly accepted as a legitimate transaction.  FN: the legitimate transaction instrument-terminal pair is rejected incorrectly.
\end{itemize}

In order to determine an optimal similarity threshold to balance the security and usability of each sensor, 100 thresholds -- between the minimum and maximum observed distances for MAE and $[-1,1]$ for correlation -- were tested for each similarity metric.  The FPR and FNR were measured at each threshold using Eqs. \ref{eq:tpr} and \ref{eq:fpr}.  Here, a threshold is the maximum permitted difference in similarity before a transaction is rejected; conversely, a transaction is accepted if the similarity in device sensor measurements is within this threshold.  Ideally, a chosen threshold for a real-world deployment should reject all illegitimate transactions, namely those between the distant instrument and the terminal, while accepting all legitimate transactions between the transaction instrument and terminal.   In our analysis, the EERs and associated thresholds for each sensor are calculated for both of the similarity metrics described in Section \ref{sec:sim}.

\begin{equation}
	TPR = \frac{TP}{TP + FN}
	\qquad
	TNR = \frac{TN}{TN + FP}
	\label{eq:tpr}
\end{equation}
\begin{equation}
	FPR = 1 - TNR 
	\qquad
	FNR = 1 - TPR
	\label{eq:fpr}
\end{equation}

\begin{figure}[ht]
	\centering
	\includegraphics[width=0.57\columnwidth]{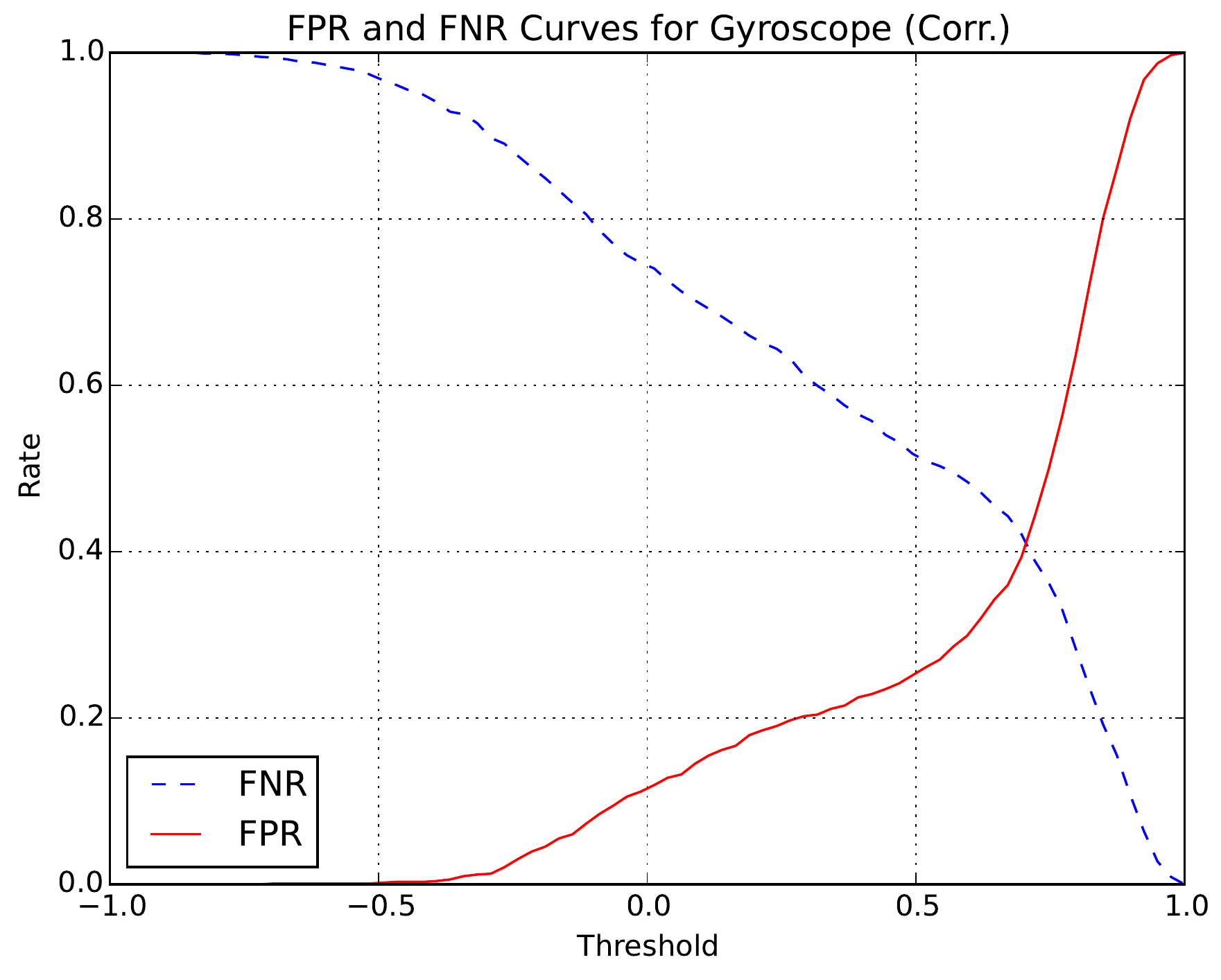}
	\caption{Example Gyroscope FPR-FNR curves using correlation for Evaluation 2.}
	\label{fig:eer}
\end{figure}

Figure \ref{fig:eer} shows the FPR/FNR graph produced for 100 different thresholds using the Gyroscope sensor and Pearson's Correlation Coefficient.  By inspection, the point at which FNR and FPR intersect (FNR=FPR) has a corresponding EER of approximately 42\% for this sensor and similarity metric.

\subsection{Individual Sensor Results}

In this section, we present the number of TPs, FPs and EERs for each sensor and for both similarity metrics described in Section \ref{sec:sim}.  The results of Evaluation 1 -- the rate at which the transaction instrument and terminal pairs can be correctly identified from sensor data -- are presented below in Table \ref{tab:ev1}.

\begin{table}[ht]
	\centering
	\caption{Optimum Thresholds and EERs for Evaluation 1}
	\label{tab:ev1}
		\begin{tabular}{@{}lcccc@{}}
			\toprule
			\multirow{2}{*}{\textbf{Sensors}} & \textbf{Optimum} & \multirow{2}{*}{\textbf{$EER_{MAE}$}} & \textbf{Optimum} & \multirow{2}{*}{\textbf{$EER_{corr}$}} \\ 
			& \textbf{Threshold$_{MAE}$} & & \textbf{Threshold$_{corr}$}& \\
			\midrule
			\textbf{Accelerometer} & 0.607 & 0.650 & 0.020 & 0.591 \\
			\textbf{Gyroscope} & 0.369 & 0.551 & 0.510 & 0.520 \\
			\textbf{Magnetic Field} & 88.07 & 0.505 & 0.020 & 0.511 \\
			\textbf{Rotation Vector} & 3.02e-08 & 0.587 & 0.061 & 0.508 \\
			\textbf{Gravity} & 8.57-07 & 0.507 & 0.061 & 0.501 \\
			\textbf{Light} & 230.9 & 0.583 & 0.020 & 0.514 \\
			\textbf{Linear Acceleration} & 0.589 & 0.513 & -0.020 & 0.572 \\
			\bottomrule
		\end{tabular}
\end{table}

Next, for Evaluation 2, new thresholds are generated that now consider the distant instrument--terminal pair, outlined in the beginning of this section. The results -- the rate at which distant instrument and legitimate transactions are accepted and denied -- are presented in Table \ref{tab:ev2}.  A more detailed breakdown of the absolute number of TPs, TNs, FPs and FNs of Evaluation 2 is presented in Table \ref{tab:ev22}, which are generated using the EER threshold.  The rate of accepted transactions using the distant instrument -- analogous to the rate of successful relay attacks -- can be found by reading the EER directly.  By definition, the EER is the point at which FPR=FNR, and FPR are equivalent to the proportion of successful distant transactions (FPs) to the total of FPs and correctly denied distant transactions (TNs).  Hence, the rate of potentially successful relay attacks (FPR) is equal to the EER, which we present in Table \ref{tab:ev2}.  We estimate, for example, that using the accelerometer with MAE will result in 55.8\% relayed transactions being accepted.

\begin{table}[ht]
	\centering
	\caption{Optimum Thresholds and EERs for Evaluation 2}
	\label{tab:ev2}
		\begin{tabular}{@{}lcccc@{}}
			\toprule
			\multirow{2}{*}{\textbf{Sensors}} & \textbf{Optimum} & \multirow{2}{*}{\textbf{$EER_{MAE}$}} & \textbf{Optimum} & \multirow{2}{*}{\textbf{$EER_{corr}$}} \\ 
			& \textbf{Threshold$_{MAE}$} & & \textbf{Threshold$_{corr}$}& \\
			\midrule
			\textbf{Accelerometer} & 0.509 & 0.558 & 0.013 & 0.523 \\
			\textbf{Gyroscope} & 0.166 & 0.776 & 0.722 & 0.444 \\
			\textbf{Magnetic Field} & 88.12 & 0.389 & 0.038 & 0.512 \\
			\textbf{Rotation Vector} & 2.961 & 0.522 & 0.063 & 0.547 \\
			\textbf{Gravity} & 1.00e-06 & 0.430 & 0.089 & 0.492 \\
			\textbf{Light} & 25.13 & 0.739 & 0.089 & 0.492 \\
			\textbf{Linear Acceleration} & 0.516 & 0.632 & -0.013 & 0.453 \\
			\bottomrule
		\end{tabular}

\end{table}

\begin{table}[ht]
	\centering
	\caption{Breakdown of TPs, TNs, FPs and FNs for Evaluation 2}
	\label{tab:ev22}
        \begin{threeparttable}
          \begin{tabular}{@{}llcccclcccc@{}}
              \toprule
              \textbf{Sensor} & \emph{MAE} & \textbf{TPs} & \textbf{TNs} & \textbf{FPs} & \textbf{FNs} & \emph{Corr} & \textbf{TPs} & \textbf{TNs} & \textbf{FPs} & \textbf{FNs}\\ \midrule
              \textbf{Accelerometer}&  & 497 & 406 & 512 & 421 & & 539 & 438 & 480 & 379 \\
              \textbf{Gyroscope} & & 230 & 227 & 787 & 784 & & 620 & 564 & 450 & 394 \\
              \textbf{Magnetic Field}& & 630 & 618 & 390 & 378 & & 511 & 492 & 516 & 497  \\
              \textbf{Rotation Vector}& & 534 & 486 & 530 & 482  & & 509 & 460 & 556 & 507  \\
              \textbf{Gravity}& & 757 & 579 & 437 & 259 & & 593 & 516 & 500 & 423 \\
              \textbf{Light}& & 310 & 267 & 755 & 712 & & 560 & 547 & 475 & 462 \\
              \textbf{Linear Acceleration}& & 438 & 375 & 644 & 581 & & 598 & 557 & 462 & 421 \\
              \bottomrule
          \end{tabular}
        \end{threeparttable}
\end{table}

\subsection{Analysis Equipment}

Both analyses were conducted on an Ubuntu-based machine equipped with an eight-core Intel Xeon E5-2407 CPU (2.20GHz) and 8GB RAM.  The data analysis applications were developed in Python, using the NumPy \cite{numpy} library for general numerical computation and handling, and SciPy \cite{scipy} for implementations of Pearson's Correlation Coefficient and linear interpolation.  Overall, Evaluation 1 took a total of 9.5 hours to complete, including the pre-computation of $8000 \times 8000$ transactions for each similarity metric, while Evaluation 2 was performed in 304.5 seconds -- both of which were timed within the Python scripts.

\section{Outcome and Future Directions}
\label{sec:OutcomeandFutureDirections}

The results we present provide an empirical foundation for evaluating various mobile sensors as proximity detection and anti-relay mechanisms for NFC-based mobile transactions. As discussed in the methodology of Evaluation 1, the higher the EER, the greater the likelihood that a genuine transaction is rejected when readings between the terminal and transaction instrument are considered.  Next, based on Evaluation 2, we investigated the success rate of a potential relay attack by considering readings between the terminal--distant instrument and the terminal--transaction instrument. The EER columns in Table \ref{tab:ev2} presents the potential success rate of a relay attack.  Here, it can be observed that the Magnetic Field sensor performs relatively better with 38.9\% relay attack success.  This is compared with the EER for the Gyroscope sensor, which has a 77.6\% possibility of a successful relay attack taking place -- one of the highest in our analysis.  However, not only does the EER denote the potential success of a relay attack, but also the potential of a legitimate transactions being denied.  The aforementioned Magnetic Field EER indicates that 38.9\% of relay attacks would be accepted \emph{and} 38.9\% legitimate transactions would be denied.  While a high number, the rejection rate for legitimate transactions for the same sensor without DTI was 50.5\% (Evaluation 1, Table \ref{tab:ev1}).  Therefore proximity detection improved when taking relay attack data into account; this was the case for all other sensors except Gyroscope, Light and Linear Acceleration.

Nevertheless, the best performing sensor (Magnetic Field with MAE) still had a significantly high EER (0.389).  Denying over 1-in-3 legitimate transactions would invariably cause annoyance issues for users in practice.  Hence, based on our results, it is difficult to recommend any of the sensors for a single tap-single sensor deployment for high security applications, such as banking.  These sensors, however, might be appropriate for low-security access control, but we recommend that a thorough analysis of the sensors and their performance in the chosen domain is performed prior to deployment.

One potential reason that related research (discussed in section \ref{sec:RelatedWork}) in this domain has achieved different results is due to the larger sampling duration and limited field trials in other work.  The sample duration limit imposed during our experiments was in line with the performance requirements of an EMV application, i.e.\ 500 milliseconds.  Additionally, transportation is one of the biggest application areas for contactless smart cards, along with banking; in this domain, the recommended duration for  transactions is far lower, in the range of 300-400 milliseconds.  Imposing a limit of 500 millisecond in our experiments, therefore, was based on the upper bound of the recommendations of two significant application areas where contactless mobile phones might be utilised. 

As part of our ongoing research, we are currently experimenting with:

\begin{itemize}
\item Can we simultaneously measure multiple sensors within the transaction time duration?  If so, will it reduce the risk associated with these sensors individually?
\item Combining sensor measurements with time slices: only one sensor is measured at a time, but over the duration of the transaction multiple sensors could be used.
\item Is a location-specific deployment feasible? And will it improve the success rate significantly if such an approach is adopted?
\end{itemize}


\subsection{Related Work}
\label{sec:RelatedWork}
As noted previously, distance bounding protocols might not be suitable for mobile-based contactless transactions.  In existing work, therefore, ambient sensors have been proposed as a strong proximity detection mechanism to counter relay attacks, which is discussed below.

Drimer et al.\ \cite{DrimerM07} and Ma et al.\ \cite{6378376} showed how location-related data, using a GPS (Global Positioning System), can be used to determine the proximity of two NFC mobile phones.  Ma et al.\ used a ten second window with location information collected every second, which was subsequently compared across various devices.  The authors reported a high success rate in identifying the devices in close proximity to one another.

Halevi et al.\ \cite{Halevi2012} demonstrated the suitability of using ambient sound and light for proximity detection.  Here, the authors analysed the sensor measurements -- collected for 2 and 30 seconds duration for light and audio respectively -- using a range of similarity comparison algorithms.  Extensive experiments were performed in different physical locations, with a high success rate in detecting co-located devices.

Varshavsky et al.\ \cite{Varshavsky2007} based their proximity detection mechanism on the shared radio environment of devices -- the presence of WiFi access points and associated signal strengths -- using the application scenario of secure device pairing.  In this work, they considered this approach to produce low error rates, recommending it as a proximity detection mechanism.  While their paper did not focus on NFC-based mobile transactions, their techniques and methodology may still be applicable.

  \begin{table}[t]
	\centering
	\caption{Related Work in Sensors as Anti-Relay Mechanism}
	\label{tab:RelatedWork}

		\begin{tabular}{@{}lccc@{}}
		\toprule
        \textbf{Paper} & \textbf{Sensor(s) Used} & \textbf{Sample Duration} & \textbf{Contactless Suitability} \\
        \midrule
        Ma et al.\ \cite{6378376} & GPS & 10 seconds & Unlikely \\ 
         Halevi et al.\ \cite{Halevi2012} & Audio & 30 seconds & Unlikely \\ 
         								  & Light & 2 seconds & More Likely \\ 
         Varshavsky et al.\ \cite{Varshavsky2007} & WiFi (Radio Waves) & 1 second & More Likely \\
         Urien et al.\ \cite{Urien201428} & Temperature & N/A & - \\ 
         Mehrnezhad et al.\ \cite{mehrnezhad2014tap} & Accelerometer & 0.6 to 1.5 Seconds & More Likely\\
         Thruong et al.\ \cite{truong2014comparing} & GPS Raw Data & 120 seconds & Unlikely \\
         											& Wifi		   &  30 seconds & Unlikely \\
                                                    & Ambient Audio & 10 seconds & Unlikely \\
                                                    & Bluetooth & 12 seconds & Unlikely \\
        Shrestha et al. \ \cite{shrestha2014drone} & Temparture (T) & Few seconds & Unlikely \\
        										   & Precision Gas (G) & Few seconds & Unlikely \\
                                                   & Humidity (H) & Few seconds & Unlikely \\
                                                   & Altitude (A) & Few seconds & Unlikely \\
                                                   & HA	& Few seconds & Unlikely \\
                                                   & HGA & Few seconds & Unlikely \\
                                                   & THGA & Few seconds & Unlikely \\
		\bottomrule
		\end{tabular}
\end{table}

Urien et al.\ \cite{Urien201428} used ambient temperature with an elliptic curve-based RFID and/or NFC authentication protocol to determine whether two devices were co-located.  Using this method, they were successful in establishing a secure channel; the proposal combines the timing channels in RFID, traditionally used in distance bounding protocols, in conjunction with ambient temperature.  Their proposal, however, was not implemented and so there is no experimental data provided to evaluate its efficacy.

Mehrnezhad et al.\ \cite{mehrnezhad2014tap} proposed the use of an accelerometer to provide assurance that the mobile phone is within the vicinity of the payment terminal.  Their proposal requires the user to tap the payment terminal twice in succession, after which the sensor streams of the device and the payment terminal are compared for similarity.  It is difficult to deduce the total time it took to complete one transaction in its entirety, but the authors have provided a sensor recording time range of 0.6--1.5 seconds.

Thruong et al.\ \cite{truong2014comparing} evaluated four different sensors. Similarly to previous studies, their sample rates were 10-120 seconds. Although results were positive, the sample duration made them unsuitable for NFC-based mobile transactions. Shrestha et al.\ \cite{shrestha2014drone} used specialised hardware known as Sensordrone, with a number of ambient sensors, but did not evaluate the commodity ambient sensors available on commercial handsets, did not provide the sample duration, and only mentioned that data from each sensor was collected for a few seconds. This potentially renders the technique inapplicable to NFC-based mobile transactions.

We summarise the related work in Table \ref{tab:RelatedWork}, and use sensor sampling durations to determine whether a given approach is suitable for contactless NFC mobile phone transactions, namely in banking and transportation.  \emph{`Possibly not'} are those proposals whose sample duration is so large that they may not be adequate for mobile-based services that substitute contactless cards, whereas those whose sample range can be considered reasonable in certain applications are labelled as \emph{`More Likely'} in the Table \ref{tab:RelatedWork}. However, even schemes denoted as \emph{`More Likely'} may not be suitable for certain domains, such as banking or transport applications, where a strict upper-bound is often present in which to complete the entire transaction.  In these domains, the goal is to serve people as quickly as possible to maximise customer throughput, so time is critical in determining whether a transaction is successful and, indeed, permitted.  An optimum transaction duration is 500 milliseconds, rather than seconds as discussed in previous sections. 


The use of ambient sensors for proximity detection in NFC-based mobile services is expanding, as illustrated by the number of proposals that currently exist.  In this paper we extended the discussion to a large set of ambient sensors and included real-world relay attack data -- not only analysing their effectiveness as proximity detection mechanism but also as a strong anti-relay mechanism. Table \ref{tab:availability} shows that we have undertaken a comprehensive evaluation of ambient sensors for proximity detection and anti-relay effectiveness. A point to note is that in previous literature, ambient sensors were not evaluated for their effectiveness as a anti-relay mechanism (Table \ref{tab:RelatedWork}). Without a strong ability to detect relay attacks, the proximity detection alone does not warrant their deployment as an effective mechanism in NFC based mobile transactions.   


\section{Conclusion}
The aim of this investigation was to evaluate and analyse a range of sensors present in modern day off-the-shelf mobile devices, and to determine which sensors, if any, would be suitable as proximity detection and anti-relay mechanisms in the domain of NFC mobile phone transactions.  We shortlisted 17 sensors accessible through the Google Android platform, before limiting the investigation to those which were widely available and shown promising results in our initial trails.  In the existing literature, only 12 sensors have been proposed as effective proximity detection mechanisms, as listed in Table 5. Some of these sensors are only available in specialised ambient sensor hardware and in almost all instances no relay attack data was collected to determine their effectiveness against such attacks. In this study, we implemented and evaluated 7 sensors and collected data representing a genuine transactions and a malicious transaction (from actual relay attacks). The objective of collecting these two separate set of data at the same time was to empirical evaluate the implemented sensors with high degree of objectivity. We have also shown that detection rate improves for legitimate transactions if the optimum thresholds are generated from data including relay attack data -- however, even with this improvement the overall results were not promising. The data collection was carried out with the help of student volunteers at four different locations on university campus with varied ambient conditions. 
The scope of our analysis focused on NFC-enabled mobile devices that emulate traditional smart card services, such as transportation and banking. Within this scope, we have shown that non of the evaluated sensors at their current stage are effective against relay attacks -- a major point used to justify their application by existing literature. 

We will make the source code for our test-bed publicly available, along with our collected data sets, for open scrutiny and further analysis.

\bibliographystyle{acm}
\bibliography{paper}

\appendix
\section{Ambient Sensors}
\label{sec:AmbientSensors}
This appendix provides a short description of each sensor. 

\subsection{Accelerometer}
The accelerometer sensor -- deployed in most modern smartphones -- measures the acceleration applied to the device on the $x$, $y$ and $z$ axes; its units are metres per second per second ($ms^{-2}$).

\subsection{Ambient Temperature}
The ambient temperature sensor returns the room temperature in Celsius degrees ($^{\circ}$C).

\subsection{Bluetooth}
Bluetooth is a technology that facilitates wireless communication and operates in the ISM band centred at 2.4 gigahertz.  As a proximity sensor, we measure the Bluetooth devices in the vicinity (their names and MAC addresses).

\subsection{Geomagnetic Rotation Vector (GRV)}
The GRV sensor measures the rotation of the device using the device's magnetometer and accelerometer; it returns a vector containing the angles that the device is rotated in the $x$, $y$ and $z$ axes.

\subsection{Global Positioning System (GPS)}
The GPS sensor based a satellite-based global positioning and velocity measurement. A latitude and longitude pair is returned, representing a geographical location on Earth.

\subsection{Gravity}
The gravity sensor on mobile handsets measures the effect of Earth's gravity on the device, measured in metres per second per second ($ms^{-2}$).

\subsection{Gyroscope}
The gyroscope measures the rate of rotation of the device about the $x$, $y$ and $z$ axes; its units are radians per second ($rads^{-1}$).

\subsection{Relative Humidity}
The relative humidity sensor returns the percentage of the relative ambient humidity in the air.

\subsection{Light}
The light sensor measures the lighting conditions surrounding the mobile handset.  Android measures this quantity in $lux$.

\subsection{Linear Acceleration}
The linear acceleration sensor measures the affect of a device's movement on itself; its units are metres per second per second ($ms^{-2}$).

\subsection{Magnetic Field}
The magnetic field sensor detects the Earth's magnetic field along three perpendicular axes $x$, $y$ and $z$.  Android measures these values in microteslas ($\mu T$).

\subsection{Network Location}
A latitude and longitude pair is returned, representing a geographical location on Earth.

\subsection{Pressure}
The pressure sensor measures the atmospheric pressure surrounding the mobile handset. It is measured in hectopascals ($hPa$).

\subsection{Proximity}
The proximity sensors detects distance, measured in centimetres.
In many devices the sensor returns only a boolean value, declaring whether something is in close proximity to the device or not.

\subsection{Rotation Vector}
Rotation vector is a software sensor, similar to the GRV, but also incorporates the gyroscope. The returned values represent the angles which the device has rotated through the $x$, $y$ and $z$ axes.

\subsection{Sound}
For the sound sensor's measurement, we use the device's microphone to record the noise in the vicinity of the mobile handsets and retrieve the maximum amplitude that was sampled, every time it becomes available by the Android operating system.

\subsection{WiFi}
This sensor uses traditional WiFi to detect the networks in the vicinity of the mobile device.  The MAC addresses and ESSIDs of the nearby networks are returned.

\end{document}